\documentclass[showpacs,preprintnumbers,amsmath,amssymb,aps, prd]{revtex4}
\usepackage{graphicx}
\usepackage{natbib}
\usepackage{grffile}
\usepackage{caption}
\usepackage{subcaption}
\usepackage{float}
\usepackage{color}
\usepackage{dcolumn}
\usepackage{bm}
\begin{document}
\title{Chiral $QED_2$ with Faddeevian anomaly in the context of the augmented superfield approach}
\author{Sanjib Ghoshal}
\affiliation{Hooghly Mohsin College, Chinsurah, Hooghly-712101
West Bengal, India}
\author{Anisur Rahaman}
 \email{anisur.associates@aucaa.ac.in;
 manisurn@gmail.com (Corresponding Author)}
\affiliation{Durgapur Government College, Durgapur, Burdwan -
713214, West Bengal, India}

\date{\today}
\begin{abstract}
\begin{center}
Abstract
\end{center}
We consider the bosonized version of the Chiral Schwinger model in
$(1+1)$ dimension with the generalized Faddeevian anomaly, which
does not have the Lorentz covariance structure and does not have
gauge invariance either. BRST embedding is made possible after
making it gauge invariant by the incorporation of Wess-Zumino
field. For this $(1+1)$ dimensional anomalous model, we use the
Bonora-Tonin superfield formalism to construct the nilpotent and
absolutely anti-commuting anti-BRST as well as anti-co-BRST
symmetry transformations. We use the gauge-invariant constraints
on the superfields defined onto the (2, 2)-Dimensional
supermanifold along with the dual horizontality criteria.  We
provide the conserved charges linked to the aforementioned
nilpotent symmetries as well as their geometric interpretation.
The anti-BRST and anti-co-BRST charges' nilpotency and total
anticommutativity. It has also been confirmed that, in the context
of the augmented superfield formalism, the anti-BRST and
anti-co-BRST charges are nilpotent and absolutely
anti-commutative. One notable aspect of the current study is the
application of the dual-horizontality requirement to obtain
appropriate anti-co-BRST symmetry.

\end{abstract}
\maketitle
\section{Introduction}
Hagen \cite{HAG} formulated the Chiral Schwingermodel, which
replaced the vector interaction of the renowned Schwinger model
\cite{SCH, LOW} with chiral interaction. The non-unitarity issue
made the model suffer for a very long time. When Jackiw and
Rajaraman introduced a one-parameter class of anomaly ingeniously
into the model, they were able to save it from that long-term
suffering \cite{JR}, which garnered significant attention for the
model. The model takes on a physically reasonable structure even
when classical gauge symmetry breaks down drastically at the
quantum level. The chiral Schwinger model is referred to as the
Jackiw-Rajaraman variant. On the language of Dirac \cite{DIR}
Numerous approaches have been taken to carry out extensive
research on this model Similar to the Schwinger model \cite{ROT1,
ROT2, FALC, MIT, PMSG, KH, WOT, MR1, MR2, ARPL, ARSG, ARN, ARAN,
ARNP}. The conversion of this model into the gauged model of a
chiral boson \cite{SIG, FJ} with the imposition of a chiral
constraint in the phase space is an intriguing feature of this
model that resulted from the work of Harada \cite{KH}. Similar to
the Schwinger model, this model also can explain mass generation
through the use of a type of dynamical symmetry breaking. The
conversion of this model into the gauged model of a chiral boson
\cite{SIG, FJ} with the imposition of a chiral constraint in the
phase space is an intriguing feature of this model that resulted
from the work of Harada \cite{KH}. Eventually, Mitra demonstrated
that the model also developed a physically feasible structure
using a parameter-free anomaly \cite{MIT}. Mitra's version was not
manifestly Lorentz covariant, in contrast to the Jackie-Rajaraman
version. Nevertheless, the model maintained its physical
plausibility, was perfectly solvable, and produced a theoretical
spectrum that was Lorentz invariant. The anomaly in this instance
was a member of the Faddeevian category \cite{FADDEEV, FADDEEV1}.
A one-parameter class of Faddeevian anomaly was demonstrated to
make the chiral Schwinger model physically plausible in
\cite{WOT}. Similar to the chiral Schwinger model proposed by
Jackiw-Rajaraman, this model can be characterized in terms of a
chiral boson.

Furthermore, it is an anomalous model. Because of this, gauge
symmetry is violated at the quantum mechanical level, making it a
second-class constrained system as well. This research on gauge
and BRST symmetry would be interesting and instructive for this
model. According to \cite {SUDHA}, this lower-dimensional system
provides physical realizations of the de Rham cohomological
operators of differential geometry in the language of its symmetry
properties and the conserved charges corresponding to the
symmetry, making it a precise example of the Hodge theory
\cite{HODGEA, HODGEB, HODGEC, HODGED}.  Our present endeavor in
this study is to compute the proper nilpotent anti-BRST and
anti-co-BRST symmetry transformations of the enhanced
BT-superfield formalism \cite{LBON1, LBON2}, which together lead
to the derivation of a singular bosonic symmetry transformation
\cite{SUDHA}. The cohomological operators are physically realized
as a result of these symmetries taken together. One of the
geometrically intuitive ways to comprehend the abstract
mathematical features associated with the appropriate anti-BRST
symmetries in the context of geometrical aspects on the
supermanifold is through the aforementioned superfield approach to
BRST formalism \cite{LBON1, LBON2, RDEL1, RDEL2, NAKANISHI, RPM1,
RPM2, RPM3}. This superfield technique is commonly used to
generalize a given D-dimensional ordinary gauge theory onto a (D,
2)--Dimensional supermanifold, where the superfields are specified
as the dynamic fields of the supplied gauge theory. The (D,
2)-dimensional superspace coordinates  $Z^M = (x^\kappa, \theta,
\bar{\theta}$  define the (D, 2)-dimensional supermanifold. The
anti-BRST symmetry transformations of the fields specified within
the extended space (superspace) are derived from the
gauge-invariant conditions of a theory. On the other hand, the
translational generators  $(\partial_\theta,
\partial_{\bar{\theta}})$ along the Grassmanian directions
$(\theta, \bar{\theta})$ serve as the geometric basis for the
conserved charges linked to the anti-BRST symmetry. The
supermanifold $ Z^M = (x^\kappa, \theta, \bar\theta)$ is defined
as follows: the pair of variables $(\theta, \bar{\theta})$
represents the Grassmannian variables with the properties
$\theta^2= \bar{\theta}^2=0, \theta \bar{\theta}
+\bar{\theta}\theta=0$, and the bosonic variables of the given
D-dimensional ordinary gauge theory are denoted by $x^\kappa$
(with $\kappa = 0, 1, 2...D-1$).

\section{Description of the gauged model of chiral boson}
The vector Schwinger model in (1+1)-dimensional electrodynamics
refers to a model where the vector interaction between the fermion
and the gauge field is considered. According to what has been
reported, Hagen \cite{HAG} developed the chiral Schwinger model by
switching from vector-type interaction to chiral type. The
generating functional of the chiral Schwinger model is provided by
\begin{equation}
{\cal Z}= \int d\Psi d\bar{\Psi}d{\cal A}_\mu e^{\int d^2x{\cal
L}_{FCS}}
\end{equation}
where
\begin{equation}
{\cal L}_{FCS}=\bar{\Psi}(i\gamma^{\mu}\partial_{\mu} +
e(1+\gamma_5)\gamma^{\mu} A_{\mu} )\Psi-\frac{1}{4} F_{\mu\nu}
F^{\mu\nu}.
\end{equation}
The matter field and gauge field in this case are denoted by $Psi$
and $cal A_mu$, respectively. In (1+1) dimension, the values $0$
and $1$ are taken by the indices $\mu, \nu$. The absence of the
unitarity condition caused the model to suffer for a very long
period. The long-suffering of this model was eliminated by Jackiw
and Rajaraman \cite{JR} inviting anomaly into the picture. The
model with Jackiw-Rajaraman type anomaly in its bosonized form is
given by
\begin{eqnarray}
L &=& \int dx[
\frac{1}{2}(\dot{\phi}^{2}-\phi'^{2})+e(\dot{\phi}+\phi')({
A}_{0}-{ A}_{1}) +\frac{1}{2}(\dot{{ A}_1}-{ A}'_{0})^{2} +
\frac{1}{2}e^{2}A_{\mu} A^{\mu}
\end{eqnarray}
Here $\phi$ represents a boson field.  The bosonized version of
the chiral Schwinger model with a generalized Lorentz
non-covariant masslike term \cite{WOT} is given by
\begin{eqnarray}
L_{BCS}= \int dx[
\frac{1}{2}(\dot{\phi}^{2}-\phi'^{2})+e(\dot{\phi}+\phi')({
A}_{0}-{ A}_{1}) +\frac{1}{2}(\dot{{ A}_1}-{ A}'_{0})^{2}
+\frac{1}{2}e^{2}({ A}^2_{0} + 2\alpha { A}_{1}{ A}_{0} +
(2\alpha-1){ A}^2_{1})] \label{FADC}
\end{eqnarray}
This particular electromagnetic anomaly is a member of a unique
class called Faddeevian anomalies \cite{FADDEEV, FADDEEV1}. This
model, in contrast to the bosonized form of the Jackiw-Rajaraman
type, is not manifestly Lorentz covariant. The masslike term was
$\frac{1}{2}e^2 A_\mu A^\mu$ in the bosonized version owing to
Jackiw-Rajaraman, rather than $\frac{1}{2}e^{2}( A^2_{0}+ 2\alpha
A_{1} A_{0} + (2\alpha-1) A^2_{1})$. A massless boson and a
missive bison with squared mass $m^2= \frac{ae^2}{a-1}$ were
contained in the theoretical spectrum. On the other hand, the
model as presented in the Lagrangian density (\ref{FADC})
comprises a massless chiral boson and a massive boson; in this
case, the square of the boson's mass was
$m^2=-\frac{e^2(\alpha-1)^2}{\alpha}$. It is important to note
that before \cite{WOT} was developed, Mitra in \cite{MIT} first
presented a parameter-free masslike term related to the Faddeevian
anomaly. It is equivalent to the generalized one's $\alpha=-1$,
which was developed subsequently in \cite{WOT}. A massless chiral
boson and a large boson with mass $e^2$ are present in the
theoretical spectra of this paradigm. We need the gauge invariant
version of this model because we intend to construct the upgraded
BT-superfield formalism \cite{LBON1, LBON2, RDEL1, RDEL2,
NAKANISHI, RPM1, RPM2, RPM3} and compute its correct nilpotent
anti-BRST and anti-co-BRST symmetry transformations. There are
several ways to get a theory to be gauge invariant. We use the
Stukleberg formalism \cite{STUECK} to develop the proper
Wess-Zumino term \cite{WESS} corresponding to this model, thereby
making it gauge invariant. It is observed that the proper
Wess-Zumino term for this model is
\begin{eqnarray}
L_{WZ} &=& \int dx[
 \alpha (\dot{w} w^{'}+ {w^{'}}^{2}) -2e \alpha (A_0 +A_1)w^{'}\nonumber\\
 &-& e(\alpha +1)(A_1 \dot{w}- A_0 w^{'}).
\end{eqnarray}
Therefore, the gauge-invariant version of this model reads
\begin{eqnarray}
L_{GI} &=& \int dx[
\frac{1}{2}(\dot{\phi}^{2}-\phi'^{2})+e(\dot{\phi}+\phi')({
A}_{0}-{ A}_{1}) +\frac{1}{2}(\dot{{ A}_1}-{
A}'_{0})^{2}\nonumber\\ &+&\frac{1}{2}e^{2}({ A}^2_{0} + 2\alpha {
A}_{1}{ A}_{0} + (2\alpha-1){ A}^2_{1})]+ \alpha (\dot{w} w^{'}+
{w^{'}}^{2}) \nonumber\\ &-&2e \alpha (A_0 +A_1)w^{'}- e(\alpha
+1)(A_1 \dot{w}- A_0 w^{'})\nonumber\\ &=& L_0 + L_{WZ}.
\label{BCS}
\end{eqnarray}
We can  see it as
\begin{eqnarray}
L_{GI}= L_0 + L_{WZ}
\end{eqnarray}
where
\begin{eqnarray}
L_0 &=& \int dx[
\frac{1}{2}(\dot{\phi}^{2}-\phi'^{2})+e(\dot{\phi}+\phi')({
A}_{0}-{ A}_{1}) +\frac{1}{2}(\dot{{ A}_1}-{
A}'_{0})^{2}\nonumber\\ &+&\frac{1}{2}e^{2}({ A}^2_{0} + 2\alpha {
A}_{1}{ A}_{0} + (2\alpha-1){ A}^2_{1})]
\end{eqnarray}
The Lagrangian $L_{GI}$ is invariant under the transformation
$A_{\mu}\rightarrow A_{\mu}+\frac{1}{e}\partial_{\mu}\lambda,~~
\phi\rightarrow \phi +\lambda, ~~ w\rightarrow w-\lambda$. The
BRST invariant Lagrangian density corresponding to the Lagrangian
(\ref{BCS}) can be written down as
\begin{eqnarray}
L_{BRST} &=& L_{GI}+ \int dx[\frac{1}{2}b^2 +b
\partial_{\mu}A^{\mu}
 + \partial_{\mu}\bar{C}\partial^{\mu}C]. \label{LBRST}
\end{eqnarray}
Here $C$ and $\bar{C}$ represent ghost and anti-ghost field and
$b$ is an auxiliary field. From (\ref{LBRST}), the canonical
momenta corresponding to the fields $A_0, A_1, \phi$ and $w$ from
the standard definitions are found as follows.
\begin{equation}
\frac{\partial L}{\partial \dot{A_0}}= \pi_0 \approx 0
\end{equation}
\begin{equation}
\frac{\partial L}{\partial \dot{A_1}}= \pi_1 = \dot{A_1}- A_0^{'}
\end{equation}
\begin{equation}
\frac{\partial L}{\partial \dot{\phi}}= \pi_\phi = \dot{\phi}+
e(A_0 - A_1)
\end{equation}
\begin{equation}
\frac{\partial L}{\partial \dot{w}}= \pi_w = \alpha w^{'}- e(1+
\alpha) A_1 \label{rac}
\end{equation}
Here $\pi_0, \pi_1, \pi_\phi$, and $\pi_w$ are the  canonical
momenta corresponding to the fields $A_0, A_1, \phi$, and $w$.
Through the Legendre transformation, canonical  Hamiltonian is
then computed and that reads
\begin{eqnarray}
H_c &=& \pi_0 \dot{A_0} +\pi_1 \dot{A_1}+\pi_\phi \dot{\phi}+\pi_w
\dot{w}- L \nonumber\\ &=& \frac{1}{2}(\pi_1^{2}+ {\phi^{'}}^2+
{\pi_\phi}^2)+ \pi_1 A_0^{'}-e (A_0-A_1)(\pi_\phi
+\phi^{'})\nonumber\\ &-& [e^2 (\alpha - 1) A_1 ^{2}+ e^2
(\alpha+1)A_0A_1-e (1+\alpha)A_0 w^{'}\nonumber\\&-&\alpha
{w^{'}}^2 + 2e \alpha (A_0 +A_1)w^{'}.
\end{eqnarray}
Note that
\begin{equation}
C_1= \pi_0\approx 0 \label{PCON1}
\end{equation}
\begin{equation}
C_2= \pi_w - \alpha w^{'}+ e(1+ \alpha) A_1\approx 0 \label{PCON2}
\end{equation}
are the primary constraints of the theory. The preservation of the
constraint $\pi_0=0$  requires $\dot{\pi_0}=0$, that leads to a
secondary constraint
\begin{equation}
\pi_1^{'}+ e(\pi_\phi +\phi^{'})+e^2 (\alpha+1)A_1 +e(1-
\alpha)w^{'}\approx 0. \label{car}
\end{equation}
Therefore, the total Hamiltonian is given by
\begin{equation}
  H_T=H_c +u\pi_0+ v\lbrace\pi_w - \alpha w^{'}+ e(1+
\alpha) A_1\rbrace.
\end{equation}
Using equation (\ref{PCON2}) and (\ref{car}) we have
\begin{equation}
\pi_1^{'}+ e(\pi_\phi +\phi^{'}) +e(w^{'}- \pi_w) \approx 0.
\label{GENC}
\end{equation}
The constraints  (\ref{PCON1}) and (\ref{GENC}) forms a first
class set and work as a generator of the  gauge transformations,
from which the following transformations correspond to the fields
$A_0, A_1,\phi, w$  result.
\begin{equation}
A_{\mu}\rightarrow A_{\mu}+\frac{1}{e}\partial_{\mu}\lambda,~~
\phi\rightarrow \phi +\lambda, ~~ w\rightarrow w-\lambda.
\end{equation}
We are in a position to construct the nilpotent BRST charge.  For
this theory it reads
\begin{equation}
Q_b= iC [\pi_1^{'}+ e(\pi_\phi- \pi_w ) +e(\phi^{'}+w^{'})]-i
\pi_0 \dot{C}. \label{NILQ}
\end{equation}
The BRST transformations for the fields that follow from the BRST
charge (\ref{NILQ}) are
\begin{eqnarray}
S_b \phi &=& -eC, ~~~~~~~~~~~S_b w = eC \nonumber\\ S_b \pi_{
\phi} &=& -eC^{'},
 ~~~~~~~~~~~S_b \pi_w = eC^{'}\nonumber\\ S_b A_0 &=& \dot{C}, ~~~~~~~~~~~S_b A_1 = C^{'}
 \nonumber\\ S_b C &=& 0, ~~~~~~~~~~~S_b \bar{C} =\pi_0 =b.
\end{eqnarray}
\section{SUPERFIELD APPROACH}
This $(1+1$ dimensional bosonized version of the chiral Schwinger
model entangled with a generalized Faddeevian anomaly can be
derived from the nilpotent and absolutely anticommuting anti-BRST
symmetries using a formalism developed by Bonora-Tonin's
\cite{LBON1, LBON2}. The spacetime variables $(x, t)$ are indeed
the functions of the fields $A_0(x, t)$ and $A_1(x, t)$ in
spacetime with the dimension $(1+1$. Now let us define the
one-form connection $A(1)$ and the exterior derivative $d$.
\begin{equation}
d = dt{\partial}_t +dx {\partial}_x,
\end{equation}
\begin{equation}
A^{(1)}=dt A_0 + dx A_1.
\end{equation}
In the superfield formalism developed in \cite{LBON1, LBON2}, we
generalize the exterior derivative and one-form connection to the
super exterior derivative $\tilde{d}$ and super one-form
connection $\tilde{A}(1)$. In the superspace, there are two extra
Grassmannian variables $\theta$ and  $\bar{\theta}$  along with
the spacetime variable (x, t). The Grassmannian variables $\theta$
and b $\bar{\theta}\theta$ satisfy the relation
\begin{equation}
\theta^2=0=\bar{\theta}^2 ~~~~~~\theta\bar{\theta} +
\theta\bar{\theta}=0.
\end{equation}
\begin{equation}
\tilde{d} = dt{\partial}_t +dx {\partial}_x +d\theta
{\partial}_{\theta} +d\bar{\theta} {\partial}_{\bar{\theta}}
\end{equation}
\begin{equation}
\tilde{A^{(1)}} = dt \tilde{A_0} +dx \tilde{A_1} +d\theta \bar{F}
+d\bar{\theta} F
\end{equation}
where $\bar{A}_0$, $\bar{A}_1$ are the superfields corresponding
to $A_0$ and  $A_1$, respectively, and $F$ and $ \bar{F}$ are the
superfields corresponding to the ghost and anti-ghost field   $C$
and $\bar{C}$ respectively. Now along the Grassmannian directions
these superfields are expanded, in terms of basic and secondary
fields of the theory as follows.
\begin{equation}
\tilde{A_0} (x,t,\theta, \bar{\theta})= A_0 (x,t) + \theta
\bar{f_1} (x,t)+ \bar{\theta} f_1 (x,t) + i \theta \bar{\theta }
B_1 (x,t) \label{SA0}
\end{equation}
\begin{equation}
\tilde{A_1} (x,t,\theta, \bar{\theta})= A_1 (x,t) + \theta
\bar{f_2} (x,t)+ \bar{\theta} f_2 (x,t) + i \theta \bar{\theta }
B_2 (x,t)\label{SA1}
\end{equation}
\begin{equation}
F (x,t,\theta, \bar{\theta})= C (x,t) + i\theta b_1 (x,t)+ i
\bar{\theta} b_1 (x,t) + i \theta \bar{\theta } S_1
(x,t)\label{SF}
\end{equation}
\begin{equation}
\bar{F} (x,t,\theta, \bar{\theta})= \bar{C} (x,t) + i\theta b_2
(x,t)+ i \bar{\theta} b_2 (x,t) + i \theta \bar{\theta } S_2 (x,t)
\label{SBF}
\end{equation}
In the above expression,  $B_1, B_2, b_1, \bar{b}_1, b_2,
\bar{b}_2$ and $f_1, \bar{f}_1, f_2, \bar{f}_2, s_1, s_2$ are
introduced to promote the theory from the usual manifold to the
supermanifold. The fields $B_1, B_2, b_1, \bar{b}_1, b_2,
\bar{b}_2$ are all bosonic in character, whereas the nature of the
fields $f_1, \bar{f}_1, f_2, \bar{f}_2, s_1, s_2$ are fermionic .
Let us now implement the widely used method, the horizontality
criterion, which places the following restrictions:
\begin{eqnarray}
d A^{(1)}= \tilde{d} \tilde{A^{(1)}} \label{HC}
\end{eqnarray}
where
\begin{eqnarray}
d A^{(1)}&=& (dt {\partial}_t + dx {\partial}_x)(dt A_0+dx A_0)
\nonumber\\ &=& ({\partial}_t A_0 )dt \wedge dt + ({\partial}_t
A_1 )dt \wedge dx +({\partial}_x A_0 )dx \wedge dt \nonumber\\ &
+&({\partial}_x A_1 )dx \wedge dx \nonumber\\ &=& (\dot{A_1}-
{A_0}^{'} )dt \wedge dx
\end{eqnarray}
and
\begin{eqnarray}
\tilde{d} \tilde{A^{(1)}} &=& (dt {\partial}_t + dx {\partial}_x+
d\theta {\partial}_\theta +d {\bar{\theta}} {\partial}_
{\bar{\theta}})(dt \tilde{A_0} + dx \tilde{A_1}
+ d\theta \bar{F} +d {\bar{\theta}} F)\nonumber\\
&=& ({\partial}_t \tilde{A_0})dt\wedge dt +({\partial}_t
\tilde{A_1})dt\wedge dx+({\partial}_t \bar{F})dt\wedge d\theta
+({\partial}_t F)dt\wedge d\bar{\theta}\nonumber\\
&+&({\partial}_x \tilde{A_0})dx\wedge dt +({\partial}_x
\tilde{A_1})dx\wedge dt +({\partial}_x \bar{F})dx\wedge d\theta
+({\partial}_x F)dx\wedge d\bar{\theta}\nonumber\\
&+&({\partial}_\theta \tilde{A_0})d\theta\wedge dt
+({\partial}_\theta \tilde{A_1})d\theta\wedge dx
+({\partial}_\theta \bar{F})d\theta\wedge d\theta
 +({\partial}_\theta F)d\theta\wedge d\bar{\theta}\nonumber\\
 &+&({\partial}_{\bar{\theta}} \tilde{A_0})d {\bar{\theta}}\wedge dt
 +({\partial}_{\bar{\theta}} \tilde{A_1})d {\bar{\theta}}\wedge dx
 +({\partial}_{\bar{\theta}} \bar{F})d {\bar{\theta}}\wedge d\theta
  +({\partial}_{\bar{\theta}} F)d {\bar{\theta}}\wedge d {\bar{\theta}}\nonumber\\
  &=& (\dot{A_1}+\theta \dot{\bar{f_2}}+\bar{\theta}\dot{f_2}+i \theta \bar{\theta}\dot{B_2}
  - A_0 ^{'}-\theta \bar{f_1}^{'}-\bar{\theta}f_1 ^{'}-
   i \theta \bar{\theta}B_1 ^{'})dt\wedge dx \nonumber\\
   &+& (\dot{\bar{C}}+i\theta \dot{\bar{b_2}}+i\bar{\theta}\dot{b_2}
   +i \theta \bar{\theta}\dot{S_2}-  \bar{f_1}-i\bar{\theta}B_1)dt\wedge d \theta \nonumber\\
   &+& (\dot{{C}}+i\theta \dot{\bar{b_1}}+i\bar{\theta}\dot{b_1}+i \theta \bar{\theta}\dot{S_1}
   -  {f_1}-i{\theta}B_1)dt\wedge d \bar{\theta}\nonumber\\ &+& ({\bar{C}}^{'}
   +i\theta {\bar{b_2}}^{'}+i\bar{\theta}{b_2}^{'}+i \theta \bar{\theta}{S_2}^{'}
   -  \bar{f_2}-i\bar{\theta}B_2)dx\wedge d \theta \nonumber\\
   &+& ({{C}}^{'}+i\theta {\bar{b_1}}^{'}+i\bar{\theta}{b_1}^{'}
   +i \theta \bar{\theta}{S_1}^{'}-  {f_2}-i{\theta}B_2)dx\wedge d \bar{\theta} \nonumber\\
   &+& (i \bar{b_1}+ i \bar{\theta}S_1 - i b_2 - i \theta S_2)d\theta \wedge d \bar{\theta}
\end{eqnarray}
According to the previously given criterion, a physical quantity
must remain unchanged when Grassmannian variables are generalized
onto a (2, 2)-dimensional supermanifold. The above-mentioned
horizontality criterion {\ref{HC}} allows us to extract the
subsequent algebraic connections between the core and subsidiary
fields of the theory. By utilizing the relation {\ref{HC}}, we can
derive the following relationship between the fields.
\begin{eqnarray}
\bar{f_1} = \dot{\bar{C}},~~~~~~~~~~~ ~~~~  S_1=0,~~~~~~~~~~~~~~~~~b_1=0 \nonumber\\
f_1 = \dot{C},~~~~~~~~~~~~~~~~  S_2=0,~~~~~~~~~~~~~~~~~\bar{b_2}=0 \nonumber \\
\bar{f_2}= {\bar{C}}^{'},~~~~~~~~~~~ \bar{b_1}-b_2=0\Rightarrow
\bar{b_1}=b_2= b (say)
\nonumber\\f_2=C^{'},~~~B_1=\dot{b_2}=\dot{\bar{b_1}}= \dot{b},~~~
B_2={b_2}^{'} ={\bar{b_1}}^{'}= b^{'} \label{RH}
\end{eqnarray}
Using the relations from Eqn.(\ref{RH}), the Eqns (\ref{SA0}),
(\ref{SA1}), (\ref{SF}), and (\ref{SBF}) acquire the following
form.
\begin{eqnarray}
\tilde{A_0}^{(h)} (x,t,\theta, \bar{\theta})&=& A_0 (x,t) + \theta
\dot{\bar{C}} (x,t)+ \bar{\theta} \dot{C} (x,t)+ i \theta
\bar{\theta } \dot{b} (x,t)\nonumber\\ &=& A_0 (x,t) + \theta
(S_{ab}A_0)+ \bar{\theta} (S_b A_0) +  \theta \bar{\theta }(S_b
S_{ab} A_0)
\end{eqnarray}
\begin{eqnarray}
\tilde{A_1}^{(h)} (x,t,\theta, \bar{\theta})&=& A_1 (x,t) + \theta
{\bar{C}}^{'} (x,t)+ \bar{\theta} {C}^{'} (x,t)+ i \theta
\bar{\theta } {b}^{'} (x,t)\nonumber\\ &=& A_0 (x,t) + \theta
(S_{ab}A_1)+ \bar{\theta} (S_{b}A_1) +  \theta \bar{\theta } (S_b
S_{ab} A_0)
\end{eqnarray}
\begin{eqnarray}
{F}^{(h)} (x,t,\theta, \bar{\theta})&=& C(x,t)+i \theta
b(x,t)\nonumber\\ &=& C (x,t) + \theta (S_{ab}C)+ \bar{\theta}
(S_{b}C) +  \theta \bar{\theta } (S_b S_{ab} C)
\end{eqnarray}
\begin{eqnarray}
\bar{F}^{(h)} (x,t,\theta, \bar{\theta})&=& \bar{C}(x,t)+i \theta
\bar{b}(x,t)\nonumber\\ &=& \bar{C} (x,t) + \bar{\theta}
(S_{b}\bar{C})+ \theta (S_{ab}\bar{C}) +  \theta \bar{\theta }
(S_b S_{ab} \bar{C})
\end{eqnarray}
It is straightforward to verify that the expressions
$A_0+\frac{1}{e} \dot{\phi}$ and $A_0-\frac{1}{e} \dot{w}$ are
gauge invariant. To find the transformations of the fields $\phi$
and $w$  we promote  the gauge invariant relation to the
supermanifold which imply
\begin{equation}
A_0+\frac{1}{e} \dot{\phi}= \tilde{A_0}+\frac{1}{e}
\dot{\tilde{\phi}}\label{ge1},
\end{equation}
\begin{equation}
A_0-\frac{1}{e} \dot{w}= \tilde{A_0}-\frac{1}{e}
\dot{\tilde{w}}\label{ge2},
\end{equation}
where,
\begin{equation}
\tilde{\phi} (x,t,\theta, \bar{\theta})= \phi (x,t) +i \theta
\bar{f_3} (x,t) + i\bar{\theta} f_3 (x,t)+ i \theta \bar{\theta }
B_3 (x,t),
\end{equation}
\begin{equation}
\tilde{w} (x,t,\theta, \bar{\theta})= w (x,t) +i \theta \bar{f_4}
(x,t) + i\bar{\theta} f_4 (x,t)+ i \theta \bar{\theta } B_4 (x,t).
\end{equation}
Using condition (\ref{ge1}), we get
\begin{eqnarray}
A_0+ \frac{1}{e} \dot{\phi}&=&  A_0 (x,t) + \theta \dot{\bar{C}}
(x,t)+
 \bar{\theta} \dot{C} (x,t)+ i \theta \bar{\theta } \dot{b} (x,t)\nonumber\\
  &+&\frac{1}{e} (\dot{\phi} (x,t) +i \theta \dot{\bar{f_3}} (x,t)
  + i\bar{\theta} \dot{f_3} (x,t)+ i \theta \bar{\theta } \dot{B_3}
  (x,t)),
\end{eqnarray}
which gives
\begin{equation}
f_3= ieC,~~~~~~~~ \bar{f_3}=ie \bar{C},~~~~~~~  B_3= -eb.
\end{equation}
Similarly, the condition (\ref{ge2}) renders
\begin{equation}
f_4= -ieC,~~~~~~~~ \bar{f_4}=-ie \bar{C},~~~~~~~  B_4= eb.
\end{equation}
As a result, we have the following explicit expression of
$\tilde{\phi} (x,t,\theta, \bar{\theta})$ and $\tilde{w}
(x,t,\theta, \bar{\theta})$.
\begin{eqnarray}
\tilde{\phi} (x,t,\theta, \bar{\theta})&=& \phi (x,t) -e \theta
\bar {C} (x,t)-e\bar{\theta} C (x,t)-ie \theta \bar{\theta } b
(x,t)\nonumber\\ &=& \phi (x,t) + \theta
(S_{ab}\phi)+\bar{\theta}(S_{b}\phi) + \theta \bar{\theta } (S_b
S_{ab}\phi).
\end{eqnarray}
\begin{eqnarray}
\tilde{w} (x,t,\theta, \bar{\theta})&=& w (x,t) +e \theta \bar {C}
(x,t)+e\bar{\theta} C (x,t) +ie \theta \bar{\theta } b
(x,t)\nonumber\\ &=& w (x,t) + \theta
(S_{ab}w)+\bar{\theta}(S_{b}w) + \theta \bar{\theta } (S_b
S_{ab}w).
\end{eqnarray}
In Eqn. (\ref{NILQ}) nilpotent BRST charge is given which can be
expressed as
\begin{equation}
Q_b= \int dx(b\dot{C}-\dot{b}C),
\end{equation}
and consequently, the anti-BRST charge is
\begin{equation}
Q_{ab}= \int dx(b\dot{\bar{C}}-\dot{b}\bar{C}).
\end{equation}

\section{Dual BRST Symmetries}
We now turn our attention to the Dual BRST symmetry. To build
nilpotent and completely anticommuting anti-co-BRST symmetries, we
will employ the dual horizontality requirement and the extended
superfield approach to BRST formalism. The relation \cite{HODGED}
confines the fields by dual horizontality.
\begin{equation}
\delta A^{(1)}= \tilde{\delta} \tilde{A^{(1)}}, \label{DHOR}
\end{equation}
where $\delta$ represents the co exterior derivative $\delta=
*d*$, here $*$ represents Hodge duality operator in ordinary
spacetime, and $\tilde{\delta}=\star \tilde d \star $ is the super
co-exterior derivative, where $\star$ represents Hodge duality
operator in superspace.  The gauge-fixing term, which is invariant
under the anti-co-BRST symmetries, is implied to be unaffected
when space is extended to the super-space by amending Grassmannian
variables, according to the dual horizontality requirement
(\ref{DHOR}) mentioned above. Based on the accepted definition, we
write
\begin{eqnarray}
\star \tilde{A^{(1)}}&=& \star [dt \tilde{A_0}+ dx \tilde{A_1}+d\theta \bar{F}+ d\bar{\theta}F]\nonumber\\
&=&  (dx\wedge d\theta \wedge d \bar{\theta})\tilde{A_0}
+ (dt\wedge d\theta \wedge d \bar{\theta})\tilde{A_1}\nonumber\\
&+& (dt\wedge dx \wedge d \bar{\theta})\bar{F}+ (dt\wedge dx
\wedge d {\theta}){F},
\end{eqnarray}
and
\begin{eqnarray}
\tilde{\star \tilde{A^{(1)}}}&=& (dt\wedge dx\wedge d\theta \wedge
d \bar{\theta})\dot{\tilde{A_0}}
+ (dt\wedge dt\wedge d\theta \wedge d \bar{\theta}) \dot{\tilde{A_1}}\nonumber\\
&+& (dt\wedge dt\wedge dx \wedge d \bar{\theta})\dot{\bar{F}}
+ (dt\wedge dt\wedge dx \wedge d {\theta})\dot{F}\nonumber\\
&+& (dx\wedge dx\wedge d\theta \wedge d
\bar{\theta}){\tilde{A_0}}^{'}
+ (dx\wedge dt\wedge d\theta \wedge d \bar{\theta}) {\tilde{A_1}}^{'}\nonumber\\
 &+& (dx\wedge dt\wedge dx \wedge d \bar{\theta}){\bar{F}}^{'}
 + (dx\wedge dt\wedge dx \wedge d {\theta}){F}^{'}\nonumber\\
 &+&(d\theta\wedge dx\wedge d\theta \wedge d \bar{\theta})\partial_{\theta} {\tilde{A_0}}
  + (d\theta\wedge dt\wedge d\theta \wedge d \bar{\theta})\partial_{\theta} {\tilde{A_1}}\nonumber\\
  &-&  (d\theta\wedge dt\wedge dx \wedge d \bar{\theta})\partial_{\theta} {\tilde{F}}
  -(d\theta\wedge dt\wedge dx \wedge d \bar{\theta})\partial_{\theta} {F}\nonumber\\
  &+&(d\bar{\theta}\wedge dx\wedge d\theta \wedge d \bar{\theta})\partial_{\bar{\theta}}\bar{F}
  + (d\bar{\theta}\wedge dt\wedge d\theta \wedge d \bar{\theta})\partial_{\bar{\theta}} {\tilde{A_1}}\nonumber\\
   &-& (d\bar{\theta}\wedge dt\wedge dx \wedge d \bar{\theta})\partial_{\bar{\theta}}\bar{F}
   -(d\bar{\theta}\wedge dt\wedge dx\wedge d {\theta})\partial_{\bar{\theta}}F \nonumber\\
    &=& (dt\wedge dx\wedge d\theta \wedge d \bar{\theta})\dot{\tilde{A_0}}
    +(dx\wedge dt\wedge d\theta \wedge d \bar{\theta}) {\tilde{A_1}}^{'}\nonumber\\
     &-&  (d\theta\wedge dt\wedge dx \wedge d \bar{\theta})\partial_{\theta} {\tilde{F}}
     -(d\theta\wedge dt\wedge dx \wedge d \bar{\theta})\partial_{\theta} {F}\nonumber\\
      &-& (d\bar{\theta}\wedge dt\wedge dx \wedge d \bar{\theta})\partial_{\bar{\theta}}\bar{F}
      -(d\bar{\theta}\wedge dt\wedge dx\wedge d
      {\theta})\partial_{\bar{\theta}}F.
\end{eqnarray}
It is known that
\begin{equation}
\star (dt \wedge dx \wedge d\theta \wedge d\bar{\theta})=1.
\end{equation}
To express our result in a convenient form let us  now define
\begin{equation}
\star (dt \wedge dx \wedge d\theta \wedge d{\theta})= S^{\theta
\theta}, \label{SD1}
\end{equation}
\begin{equation}
\star (dt \wedge dx \wedge d\bar{\theta} \wedge
d\bar{\theta})=S^{\bar{\theta} \bar{\theta}}. \label{SD2}
\end{equation}
Using the above definition (\ref{SD1}, \ref{SD2}) we land onto the
exploit relation
\begin{equation}
\star \tilde{d} \star \tilde{{A^{(1)}}}= (\dot{\tilde{A_0}}-
{\tilde{A_1}}^{'} - \partial_{ \theta} \bar{F}-
\partial_{\bar{\theta}}F)-S^{\theta \theta} \partial_{ \theta} {F}
- S^{\bar{\theta} \bar{\theta}}\partial_{ \bar{\theta}} \bar{F}.
\end{equation}
If we now bring dual Horizontality condition (\ref{DHOR}) into
action it renders
\begin{equation}
\partial_{\theta}F=0\Rightarrow \bar{b_1}=0 , S_1=0,
\end{equation}
\begin{equation}
\partial_{\bar{\theta}}\bar{F}=0\Rightarrow {b_2}=0 , S_2=0
\end{equation}
\begin{equation}
\partial_{\theta}\bar{F} + \partial_{\bar{\theta}}{F}=0\Rightarrow {b_1}+ \bar{b_2}=0
\Rightarrow b_1=-\bar{b_2}=B(say).
\end{equation}
The above relation leads to the following expansion of
$F^{dh}(x,t,\theta, \bar{\theta})$ and $\bar{F} ^{dh}(x,t,\theta,
\bar{\theta})$ in the superspace.
\begin{eqnarray}
F^{dh}(x,t,\theta, \bar{\theta})&=& C(x,t)+ \bar{\theta}[i B (x,t)]\nonumber\\
 &=& C(x,t)+\theta [S_{ad}C(x,t)]+\bar{\theta} [S_{d}C(x,t)]\nonumber\\
 &+& \theta \bar{\theta}[S_dS_{ad}C(x,t)],
\end{eqnarray}
\begin{eqnarray}
&=& \bar{C}(x,t)+ {\theta}[-i B (x,t)]\nonumber\\ &=&
\bar{C}(x,t)+\theta [S_{ad}\bar{C}(x,t)]+\bar{\theta}
[S_{d}\bar{C}(x,t)]\nonumber\\ &+& \theta
\bar{\theta}[S_dS_{ad}\bar{C}(x,t)].
\end{eqnarray}
The superscript $dh$ indicates the expansion in the superspace
after employing the dual Horizontality condition (\ref{DHOR}).  We
are now able to identify the  anti-Co-BRST symmetries of the ghost
and anti-ghost fields
\begin{eqnarray}
S_d C &=& iB, ~~~~~~~~S_d C=0,~~~~~~~S_d S_{ad}C=0
\nonumber\\ S_{ad}C &=&0,~~~~~~~~ S_{ad}\bar{C}= -iB,~~~~~S_d S_{ad}\bar{C}=0\nonumber\\
 && ~~~~ S_d B= S_{ad}B=0.
\end{eqnarray}
Here $B$ is chosen to be
\begin{eqnarray}
B&=& \pi_1 ^{'}+e(\pi_{\phi}- \pi_w)+ e(\phi^{'}+w^{'})\nonumber\\
&\equiv& \bar{b} ^{'} +e(\pi_{\phi}- \pi_w)+ e(\phi^{'}+w^{'}),
\end{eqnarray}
where, $\bar{b}=\pi_1=\dot{A_1}-A_0 ^{'}$.

The Co-BRST charge has the expression
\begin{eqnarray}
Q_d&=&- \int dx \lbrace [\pi_1 ^{'}+e(\pi_{\phi}- \pi_w)+
e(\phi^{'}+w^{'})]\dot{\bar{C}}
+ \pi_0 ^{'} \bar{C} ^{'}\rbrace \nonumber\\
&=&- \int dx \lbrace [\bar{b} ^{'}+e(\pi_{\phi}- \pi_w)+
e(\phi^{'}+w^{'})]\dot{\bar{C}} - b \bar{C} ^{''}\rbrace
\nonumber\\&=&- \int dx(B\dot{\bar{C}}- \dot{B}\bar{C}).
\end{eqnarray}
Here we have used the relations $b=\pi_0$ and $b^{''}=\dot{B}$,
which are obtained from equations of motion.

The transformations for the fields are as follows.
\begin{eqnarray}
S_d \phi &=& -e\dot{\bar{C}}, ~~~~~~~~~~~S_d w = e\dot{\bar{C}}\nonumber\\
 S_d \pi_{ \phi} &=& -e\dot{\bar{C}}^{'}, ~~~~~~~~~~~
 S_d \pi_w =-e\dot{\bar{C}}^{'}\nonumber\\
 S_d A_0 &=& -\bar{C}^{''}, ~~~~~~~~~~~S_d A_1 = -\dot{\bar{C}^{'}}\nonumber\\
 S_d C &=& -i[\bar{b} ^{'}+e(\pi_{\phi}- \pi_w)+ e(\phi^{'}+w^{'})]=-iB,
 ~~~~S_d \bar{C} =0\nonumber\\ S_d b &=&0, ~~~~~~~~~~S_d \bar{b}
 =0.
\end{eqnarray}
Similarly, we have
\begin{eqnarray}
S_{ad} \phi &=& -e\dot{{C}}, ~~~~~~~~~~~S_{ad} w = e\dot{{C}}\nonumber\\
S_{ad} \pi_{ \phi} &=& -e\dot{{C}}^{'}, ~~~~~~~~~~~S_{ad} \pi_w
=-e\dot{{C}}^{'}
\nonumber\\ S_{ad} A_0 &=& -{C}^{''}, ~~~~~~~~~~~S_{ad} A_1 = -\dot{{C}^{'}}\nonumber\\
S_{ad} C &=&0, ~~~~S_{ad} \bar{C} =i[\bar{b} ^{'}+e(\pi_{\phi} -
\pi_w)+ e(\phi^{'}+w^{'})]=iB\nonumber\\ S_{ad} b &=&0,
~~~~~~~~~~S_{ad} \bar{b} =0.
\end{eqnarray}
We note that the following quantities remain invariant under
anti-Co-BRST transformation.
\begin{eqnarray}
S_d [\dot{A_0}-A_1 ^{'}]&=&0,~~~~~~~~ S_d [A_1 -\frac{1}{e} \phi^{'}]=0]\nonumber\\
S_d [\phi +w]&=&0, ~~~~~~~~~ S_d [A_1 +\frac{1}{e} w^{'}]=0].
\end{eqnarray}
As these anti-Co-BRST invariant quantities should remain
independent of Grassmannian variables we have
\begin{equation}
(\dot{\tilde{A_0}}- {\tilde{A_1}}^{'})(x,t,\theta, \bar{\theta})
=(\dot{A_0}- A_1 ^{'})(x,t,\theta, \bar{\theta})\label{con1},
\end{equation}
\begin{equation}
({\tilde{\phi}}+ {\tilde{w}})(x,t,\theta, \bar{\theta})=(\phi
+w)(x,t,\theta, \bar{\theta})\label{con2},
\end{equation}
\begin{equation}
({\tilde{A_1}}- \frac{1}{e} {\tilde{\phi}}^{'})(x,t,\theta,
\bar{\theta}) =({{A_1}}- \frac{1}{e} {{\phi}}^{'})(x,t,\theta,
\bar{\theta})\label{con3},
\end{equation}
\begin{equation}
({\tilde{A_1}}+ \frac{1}{e} {\tilde{w}}^{'})(x,t,\theta,
\bar{\theta}) =({{A_1}}+ \frac{1}{e} {{w}}^{'})(x,t,\theta,
\bar{\theta})\label{con4}.
\end{equation}
From (\ref{con1}) we find
\begin{eqnarray}
\dot{\bar{f_1}}- {\bar{f_2}}^{'}&=&0\Rightarrow
\dot{\bar{f_1}}={\bar{f_2}}^{'}\nonumber\\\dot{{f_1}}-
{{f_2}}^{'}&=&0\Rightarrow {\dot{f_1}}={{f_2}}^{'}\nonumber\\
\dot{{B_1}}- {{B_2}}^{'}&=&0\Rightarrow {\dot{B_1}}={{B_2}}^{'}.
\label{con5}
\end{eqnarray}
Consequently from (\ref{con2}) we get
\begin{eqnarray}
\bar{f_3}+\bar{f_4}&=&0\Rightarrow \bar{f_3}= -\bar{f_4}\nonumber\\
{f_3}+{f_4}&=&0\Rightarrow {f_3}= -{f_4}\nonumber
\\ {B_3}+{B_4}&=&0\Rightarrow {B_3}= -{B_4},\label{con6}
\end{eqnarray}
 (\ref{con3}) we in the same way renders
\begin{eqnarray}
\bar{f_2}-\frac{i}{e}{\bar{f_3}}^{'}&=& 0\Rightarrow \bar{f_2} =
\frac{i}{e}{\bar{f_3}}^{'}\nonumber\\ {f_2}-\frac{i}{e}{{f_3}}^{'}
&=& 0\Rightarrow {f_2}= \frac{i}{e}{{f_3}}^{'}\nonumber\\
{B_2}-\frac{1}{e}{{B_3}}^{'}&=& 0\Rightarrow {B_2}=
\frac{1}{e}{{B_3}}^{'}.\label{con7}
\end{eqnarray}
Let us now choose
\begin{equation}
f_1=-\bar{C}^{''},~~~~~~f_2= -\dot{\bar{C}^{'}},~~~~~~~~
B_1=B^{''}.
\end{equation}
Using the conditions (\ref{con5}), (\ref{con6}) and (\ref{con7})
we ultimately reach to the following conditions.
\begin{eqnarray}
\bar{f_1}&=& -C^{''},~~~~~~~ \bar{f_2}=-\dot{C^{'}},~~~~~~~B_2= \dot{B^{'}}\nonumber\\
f_3&=&-\frac{e}{i}\dot{\bar{C}}, ~~~~~~
\bar{f_3}=-\frac{e}{i}\dot{{C}},~~~~~~~ B_3= e\dot{B}\nonumber\\
f_4&=&\frac{e}{i}\dot{\bar{C}}, ~~~~~~ \bar{f_4}
=\frac{e}{i}\dot{{C}},~~~~~~~B_4= -e\dot{B},
\end{eqnarray}
\begin{eqnarray}
\tilde{A_0} (x,t,\theta, \bar{\theta})&=& A_0 (x,t) - \theta
C^{''} (x,t)-
\bar{\theta} \bar{C}^{''} (x,t)+ i\theta \bar{\theta } B^{''}(x,t)\nonumber\\
 &=& A_0 (x,t) + \theta (S_{ad}A_0)+ \bar{\theta} (S_d A_0)
 +  \theta \bar{\theta }(S_d S_{ad} A_0),
\end{eqnarray}
\begin{eqnarray}
\tilde{A_1} (x,t,\theta, \bar{\theta})&=& A_1 (x,t) - \theta
\dot{C^{'}} (x,t)
- \bar{\theta} \dot{\bar{C}^{'}} (x,t)+i \theta \bar{\theta }\dot{B^{'}} (x,t)\nonumber\\
 &=& A_1 (x,t) + \theta (S_{ad}A_1)+ \bar{\theta} (S_d A_1)
 +  \theta \bar{\theta }(S_d S_{ad} A_1).
\end{eqnarray}
\begin{eqnarray}
\tilde{\phi} (x,t,\theta, \bar{\theta})&=& \phi (x,t) - e\theta
\dot{C} (x,t)
- e\bar{\theta} \dot{\bar{C}} (x,t)+i e \theta \bar{\theta }\dot{B} (x,t)\nonumber\\
&=& \phi (x,t) + \theta (S_{ad}\phi)+ \bar{\theta} (S_d \phi) +
\theta \bar{\theta }(S_d S_{ad} \phi).
\end{eqnarray}
\begin{eqnarray}
\tilde{w} (x,t,\theta, \bar{\theta})&=& w (x,t) + e\theta \dot{C}
(x,t)
+e\bar{\theta} \dot{\bar{C}} (x,t)- i e\theta \bar{\theta } \dot{B}(x,t)\nonumber\\
 &=& w (x,t) + \theta (S_{ad}w)+ \bar{\theta} (S_d w) +
 \theta \bar{\theta }(S_d S_{ad} w).
\end{eqnarray}
\section{Nilpotency and absolute anti-commutativity in the supermanifold}
Ensuring nilpotency and anti-commutativity of BRST charges is the
natural requirement in the supermanifold like the maintenance of
these two in the usual manifold and it is instructive indeed. This
section is therefore devoted to guarantee the nilpotency and
absolute anti-commutativity property of the anti- BRST  and
anti-co-BRST charges within the framework of superfield formalism.
The BRST charge associated with the present field-theoretic model
is
\begin{equation}
Q_b= i\int dx\lbrace C  [\pi_1^{'}+ e(\pi_\phi- \pi_w )
+e(\phi^{'}+w^{'})]- \pi_0 \dot{C}\rbrace.
\end{equation}
Now we see that
\begin{equation} \pi_0 = \frac{ \partial L_{BRST}
}{\partial \dot{A_0}} = b.
\end{equation}.
Also from the equations of motion, we have
\begin{equation}
 [\pi_1^{'}+ e(\pi_\phi- \pi_w ) +e(\phi^{'}+w^{'})]= \dot{b}.
\end{equation}
So we can express the BRST charge as follows
 \begin{equation}
Q_b= i \int dx [\dot{b}C - b \dot{C}].
\end{equation}
Similarly, the anti-BRST charge can be written down as
\begin{equation}
Q_{ab}= i \int dx [\dot{b}\bar{C} - b \dot{\bar{C}}].
\end{equation}
These  charges can $Q_b$ and $Q_{ab}$ can also be expressed on the
following form by using the nilpotency property $
{{\partial}_{\bar{\theta}}}^2 =0={{\partial}_{{\theta}}}^2 $.
\begin{equation}
Q_b= i \int dx [S_b (\dot{\bar{C}}C - \bar{C} \dot{C})]= i \int dx
[S_{ab} ( - \dot{C} C)],
\end{equation}
\begin{equation}
Q_b= - \int dx [\frac{\partial}{\partial \bar{\theta}}\lbrace
{\bar{F}}^{(h)} (x,\theta, \bar{\theta}){\dot{F}}^{(h)} (x,\theta,
\bar{\theta})- {\dot{\bar{F}}}^{(h)} (x,\theta,
\bar{\theta}){{F}}^{(h)} (x,\theta,
\bar{\theta})\rbrace]_{\theta=0},
\end{equation}
\begin{equation}
Q_{ab}= i \int dx [S_{ab} (C \dot{\bar{C}} -  \dot{C}\bar{C})]= i
\int dx [S_{b} ( + {\dot{\bar{C}}} \bar{C})],
\end{equation}
\begin{equation}
Q_{ab}= + \int dx [\frac{\partial}{\partial {\theta}}\lbrace
{\bar{F}}^{(h)} (x,\theta, \bar{\theta}){\dot{F}}^{(h)} (x,\theta,
\bar{\theta})- {\dot{\bar{F}}}^{(h)} (x,\theta,
\bar{\theta}){{F}}^{(h)} (x,\theta,
\bar{\theta})\rbrace]_{\bar{\theta}=0}.
\end{equation}
Hence,
\begin{equation}
{\partial}_{\bar{\theta}} Q_b =0 ~~~~~~~ and ~~~~~~
{\partial}_{{\theta}} Q_{ab} =0.
\end{equation}
Similarly,
\begin{equation}
S_b Q_b =0 ~~~~~~~ and ~~~~~~ S_{ab} Q_{ab} =0
\end{equation}
\begin{equation}
S_b Q_b = i\lbrace Q_b, Q_b\rbrace =0 \Rightarrow {Q_b}^2 =0
\end{equation}
\begin{equation}
S_{ab} Q_{ab} = i\lbrace Q_{ab},Q_{ab}\rbrace =0 \Rightarrow
{Q_{ab}}^2 =0
\end{equation}
Also,
\begin{equation}
\lbrace Q_{b},Q_{ab}\rbrace =0
\end{equation}
The co-BRST charge associated with this theory is
\begin{eqnarray}
Q_d &=&- \int dx \lbrace [\pi_1 ^{'}+e(\pi_{\phi}- \pi_w)+
e(\phi^{'}+w^{'})]\dot{\bar{C}}+ \pi_0 ^{'} \bar{C} ^{'}\rbrace
\nonumber\\ &=&- \int dx \lbrace [\bar{b} ^{'}+e(\pi_{\phi}-
\pi_w)+ e(\phi^{'}+w^{'})]\dot{\bar{C}}- b \bar{C} ^{''}\rbrace
\nonumber\\&=&- \int dx(B\dot{\bar{C}}- \dot{B}\bar{C})
\label{QDSS}
\end{eqnarray}
In the Eqn. (\ref{QDSS}), we have used the relations $b=\pi_0$ and
$b^{''}=\dot{B}$ which are obtained from equations of motions
\begin{eqnarray}
B = \pi_1 ^{'}+e(\pi_{\phi}- \pi_w)+ e(\phi^{'}+w^{'}) \equiv
\bar{b} ^{'}+e(\pi_{\phi}- \pi_w)+ e(\phi^{'}+w^{'}),
\end{eqnarray}
where
\begin{eqnarray}
\bar{b}= \pi_1 = (\dot{A_1}- {A_0}^{'}).
\end{eqnarray}
This reveals the absolute anti-commutativity properties of
anti-BRST charges within the framework of augmented superfield
formalism.

Anti-co-BRST charge in the same fashion is given by
\begin{eqnarray}
Q_{ad}=- \int dx(B\dot{{C}}- \dot{B}{C}),
\end{eqnarray}
The charges $Q_d$ and $Q_{ad}$  can also be expressed as follows.
\begin{equation}
Q_d= -i \int dx [S_d (C \dot{\bar{C}}+ \bar{C} \dot{C})]= -i \int
dx [S_{ad} (  \dot{\bar{C}} \bar{C})],
\end{equation}
\begin{equation}
Q_d= i \int dx [\frac{\partial}{\partial \bar{\theta}}\lbrace
{\bar{F}}^{(dh)} (x,\theta, \bar{\theta}){\dot{F}}^{(dh)}
(x,\theta, \bar{\theta})- {\dot{\bar{F}}}^{(dh)} (x,\theta,
\bar{\theta}){{F}}^{(dh)} (x,\theta,
\bar{\theta})\rbrace]_{\theta=0},
\end{equation}
\begin{equation}
Q_{ad}= -i \int dx [S_{ad} ( \dot{\bar{C}}C  + \dot{C}\bar{C})]=
-i \int dx [S_{ad} ( + {\dot{{C}}} {C})],
\end{equation}
\begin{equation}
Q_{ad}= -i \int dx [\frac{\partial}{\partial {\theta}}\lbrace
{\bar{F}}^{(dh)} (x,\theta, \bar{\theta}){\dot{F}}^{(dh)}
(x,\theta, \bar{\theta})- {\dot{\bar{F}}}^{(dh)} (x,\theta,
\bar{\theta}){{F}}^{(dh)} (x,\theta,
\bar{\theta})\rbrace]_{\bar{\theta}=0}.
\end{equation}
Therefore,  it straightforward to check that
\begin{equation}
{\partial}_{\bar{\theta}} Q_d =0, ~~~ {\partial}_{{\theta}}
Q_{ad}=0.
\end{equation}

Preceding similarly as we did in the case of BRST and anti-BRST
charges, in this situation, we also have the following relations.
\begin{equation}
S_d Q_d =0, ~~~ S_{ad} Q_{ad} =0,
\end{equation}
\begin{equation}
S_d Q_d = i\lbrace Q_d, Q_d\rbrace =0 \Rightarrow {Q_d}^2 =0,
\end{equation}
\begin{equation}
S_{ad} Q_{ad} = i\lbrace Q_{ad}, Q_{ad}\rbrace =0 \Rightarrow
{Q_{ad}}^2 =0,
\end{equation}
and
\begin{equation}
\lbrace Q_{d},Q_{ad}\rbrace =0.
\end{equation}
Consequently, we have succeeded in expressing the anti-co-BRST
charges' nilpotency and absolute ant-commutativity properties in
terms of enhanced superfield formalism. The analysis leads to the
following conclusion.  Along the Grassmannian directions of the
(2, 2)-dimensional extended supermanifold, the nilpotency and
absolute anti-commutativity qualities of anti-BRST and
anti-co-BRST charges are related to the properties of
translational generators provided above.

\section{Summary and Discussion}
An informative extension to the exact derivation of the proper
nilpotent and absolutely anticommuting anti-BRST and anti-co-BRST
symmetry transformations has been performed for the (1+1)
dimensional bosonized version of the chiral Schwinger model with
Faddeevian anomaly in this endeavor, within the framework of
augmented superfield formalism. This model has a unique anomaly
structure; it falls under the Faddeevian class. Thus, gauge
invariance at the quantum level is absent. The Wess-Zumino field
is introduced to the model to achieve gauge invariance, and BRST
embedding is used to make the model conducive to the promotion of
the $(2,2)-$dimensional supermanifold from its usual $(1+1)$
dimensional manifold. In addition to the gauge-invariant
conditions, we have used the horizontality condition to construct
the anti-BRST symmetries for all the fields of the underlying
theory. Conversely, we have applied the strength of the dual-HC
condition and the anti-co-BRST invariant constraints to obtain the
entire collection of anti-co-BRST transformations for every field
in the theory that we are studying. On the $(2,2)-$dimensional
supermanifold, on which the model under consideration is promoted,
we have provided the geometrical origin of the continuous
symmetries discussed in the paper as well as the conserved charge
associated with those symmetries in terms of translation
generators along the Grassmannian directions. have also expressed
the anti-BRST and anti-co-BRST accusations in a variety of ways.
Additionally, we have demonstrated how these characteristics of
translational generators along the Grassmannian directions of the
(2, 2)-dimensional supermanifold are related to nilpotency and
absolute anti-commutativity. The utilization of the enlarged
version of superfield formalism is a novel component of our
current endeavor, which involves deriving a collection of
appropriate anti-co-BRST symmetries.

There have been multiple attempts to advance gauge theory from its
standard manifold to a higher dimensional supermanifold where
gauge symmetry is inherent in the theory. In this case, we have
undertaken the identical work for a theory in which gauge
invariance is not present. To make the theory tractable for
promotion to a supermanifold, the phasespace is thus enlarged here
and is augmented with additional superfield dimensions. This is an
innovative technique in that regard. It illustrates that after
becoming gauge invariant through the extension of the phasespace
with auxiliary fields, a theory devoid of gauge symmetry can also
be promoted to the higher dimensional supermanifold. Since these
fields remain within the non-physical domain, the auxiliary fields
do not show any negative impact on the theory.

\end{document}